\documentclass[twocolumn]{aastex63}

\submitjournal{ApJS}

\usepackage{amsmath}
\usepackage{natbib}
\usepackage{color,soul}
\usepackage{url}

\definecolor{LightCyan}{rgb}{0.88,1,1}

\newcommand{\lenstool}{{\tt{Lenstool}}}
\newcommand{\LENSTOOL}{{\tt{Lenstool}}}

\newcommand{\hst}{{\it HST}}
\newcommand{\HST}{{\it HST}}
\newcommand{\JWST}{{\it JWST}}
\newcommand{\HSTlong}{{\it Hubble Space Telescope}}

\newcommand{\relicslong}{Reionization Lensing Cluster Survey}
\newcommand{\RELICS}{RELICS}

\newcommand{\zphot}{z_{phot}}

\newcommand{\HeI}{\hbox{{\rm He}\kern 0.1em{\sc i}}}
\newcommand{\MgI}{\hbox{{\rm Mg}\kern 0.1em{\sc i}}}
\newcommand{\MgII}{\hbox{{\rm Mg}\kern 0.1em{\sc ii}}}
\newcommand{\FeII}{\hbox{{\rm Fe}\kern 0.1em{\sc ii}}}
\newcommand{\kms}{km s$^{-1}$}

\newcommand{\clustername}{SMACS\,J0723.3$-$7327}
\newcommand{\zcluster}{0.388}

\begin{document}

\title{RELICS: Strong Lens Model of SMACSJ0723.3-7327 \footnote{Based on observations made with the NASA/ESA {\it Hubble Space Telescope}, obtained at the Space Telescope Science Institute, which is operated by the Association of Universities for Research in Astronomy, Inc., under NASA contract NAS 5-26555. These observations are associated with programs GO-12166, GO-12884, GO-14096}}

\correspondingauthor{Keren Sharon}
\email{kerens@umich.edu}

\author[0000-0002-7559-0864]{Keren Sharon}
\affiliation{Department of Astronomy, University of Michigan, 1085 S. University Ave, Ann Arbor, MI 48109, USA}

\author[0000-0002-8739-3163]{Mandy C. Chen}
\affiliation{Department of Astronomy and Astrophysics, The University of Chicago, Chicago, IL 60637, USA}

\author[0000-0003-3266-2001]{Guillaume Mahler}
\affiliation{Department of Astronomy, University of Michigan, 1085 S. University Ave, Ann Arbor, MI 48109, USA}
\author[0000-0001-7410-7669]{Dan Coe}
\affiliation{Space Telescope Science Institute, 3700 San Martin Drive, Baltimore, MD 21218, USA}

\collaboration{20}{RELICS: Reionization Lensing Cluster Survey}

\begin{abstract}
We present the details of a strong lens model of SMACS J0723.3-7327, which was made public as part of the data and high level science products (HLSP) release of the RELICS \HST\ treasury program (Reionization Lensing Cluster Survey; GO-14096, PI: Coe). The model products were made available on the Mikulski Archive for Space Telescopes (MAST) via \dataset[10.17909/T9SP45]{\doi{10.17909/T9SP45}} in 2017. Here, we provide the list of constraints that were used in the \HST-based RELICS lens model, as well as other information related to our modeling choices, which were not published with the data and HLSP release. This model was computed with \LENSTOOL, used multiple images of 8 sources, with no spectroscopic redshifts. The image plane RMS was $0\farcs58$. 
\end{abstract}


\section{Introduction}
The strong lensing cluster \clustername\ was observed as one of the \JWST\ Early Release Observations (ERO) targets, and was revealed to the public in July 2022 \citep{Pontoppidan2022}.  

Following the ERO release, the field of \clustername\ was the subject of a number of publications taking advantage of JWST's capabilities, as well as the depth added by the strong lensing magnification boost, to study the background Universe, identify high-$z$ galaxies, and analyze the foreground cluster.  
In particular, the ERO prompted computation of strong lens models based on the new data \citep{mahler2022,Pascale2022,Caminha2022}.

Preceding \JWST, \clustername\ was observed by the \HSTlong\ (\hst), and its lensing signal analyzed as part of the \relicslong\ Treasury program (\RELICS; PI: Coe).
The RELICS collaboration made all high-level data products available to the community, including reduced images, catalogs, and lens models, via the Mikulski Archive for Space Telescopes (MAST) Portal at 
\dataset[10.17909/T9SP45]{\doi{10.17909/T9SP45}}\footnote{\url{https://archive.stsci.edu/prepds/relics/}}. Product-specific README files were provided, and a full description of the catalogs was given in \citet{coe2019}. However, for a large fraction of the 41 RELICS clusters, the details of the lensing analysis (multiple images, spectroscopic redshifts, and modeling choices) were not published. 
Here, we provide the community with the details of the HST-based lens model of \clustername, which was released as part of the RELICS program in 2017, in order to provide context for the public model outputs and facilitate comparisons to the new \JWST-based models. 

This modeling analysis assumed a flat cosmology with $\Omega_{\Lambda} = 0.7$, $\Omega_{m}=0.3$, and $H_0 = 70$ \kms\ Mpc$^{-1}$. In this cosmology, $1''=5.27$ kpc at the cluster redshift, $z=$\zcluster, which we rounded to $z=0.39$.

\section{Data} 
This work used \hst\ mosaics that were produced by the RELICS collaboration, combining archival imaging obtained with ACS/F606W (GO-12166) and ACS/F814W (GO-12884), with the new RELICS imaging taken with ACS/F435W, F606W, F814W, and WFC3IR/F105W, F125W, F140W, F160W (GO-14096). The depth in each filter varies between 0.5-2 orbits. For details of the observations, data reduction, photometric and photo-$z$ catalogs we refer the reader to \citet{coe2019}.  

\section{Lensing Analysis} 
\subsection{Multiple Images}
\begin{figure*}
\epsscale{1.1}
\plotone{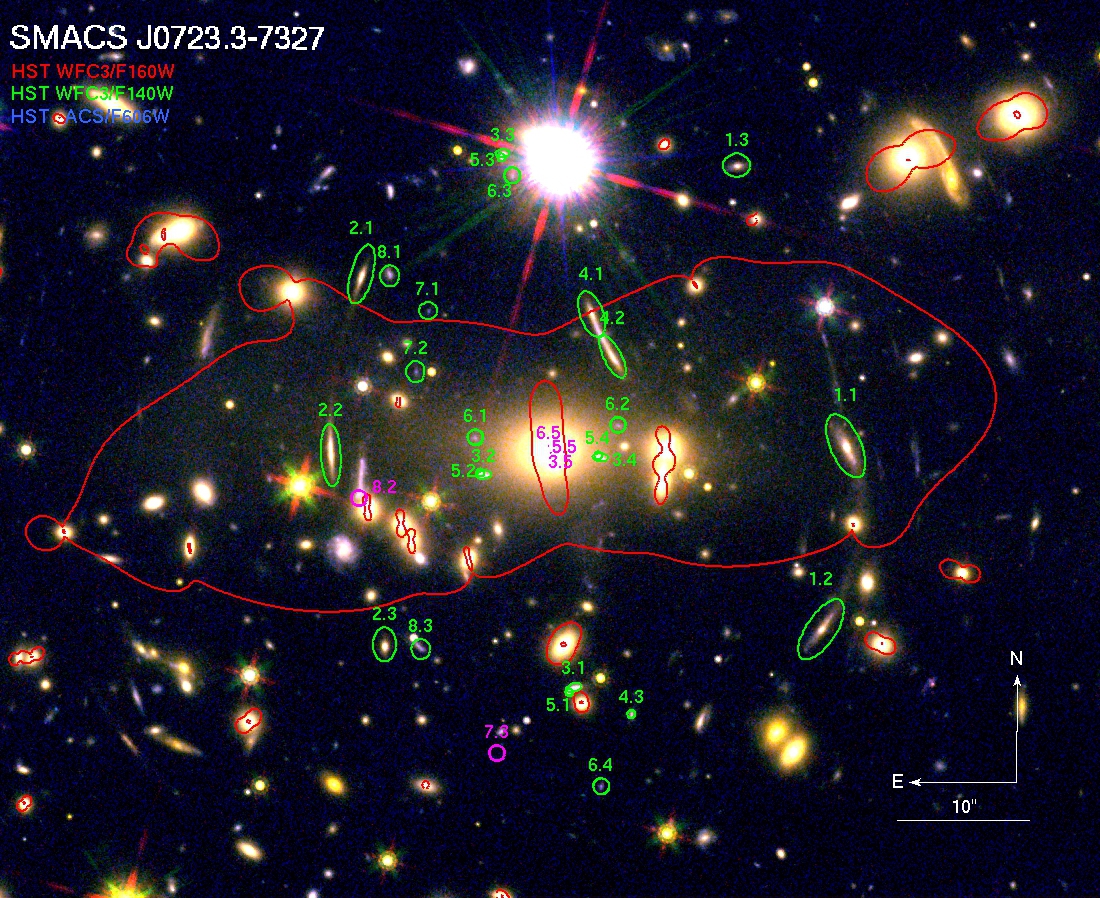}
\plotone{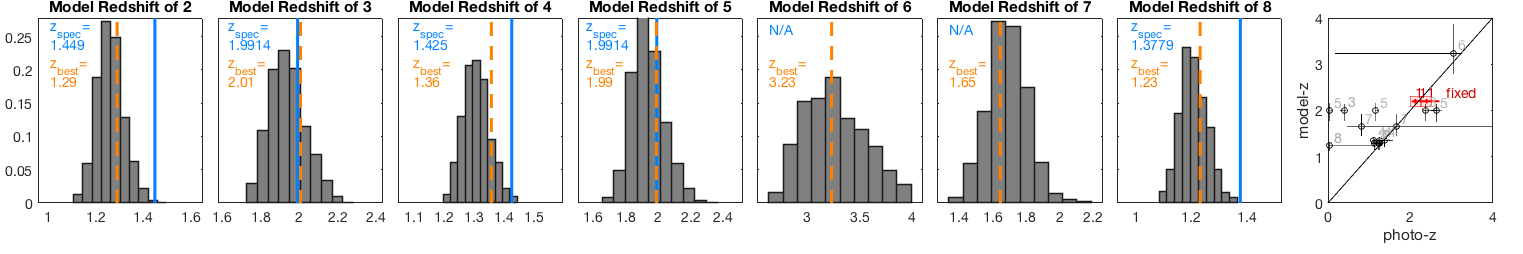}
\caption{\textit{Top:} \hst\ imaging of \clustername\ in WFC3-IR/F160W, WFC3-IR/F140W, ACS/F606W. The eight multiple image sets that were used to constrain the RELICS model are overplotted and labeled in green. Model-predicted counter images that were not used as constraints are plotted in magenta. The critical curves for a source at $z=3$ are shown in red.  \textit{Bottom:} The posterior distribution of source redshifts for sources 2--8 are shown in gray histograms (Source~1 was fixed to its photo-$z$, $\zphot=2.2$). The orange dashed line points to the best-fit model-prediction. Solid blue lines mark the recently-published spectroscopic redshifts from \citet{mahler2022}, which were not available at the time this model was computed. The rightmost panel compares the model redshift to the photometric redshifts from RELICS; this plot was used as a diagnostic tool in the modeling process. The coordinates of the lensing constraints are tabulated in \autoref{tab:arcstable}.   
} \label{fig:model}
\end{figure*}

Composite color images of the field were visually inspected to identify instances of multiple images of the same background lensed sources, to be used as lensing constraints. As is common procedure \citep[e.g.,][]{Sharon2020} we proceeded to build the lens model iteratively, starting with the most obvious and secure identification, and using preliminary iterations of the lens model to identify new constraints. 
We identified 8 sets of multiply-imaged sources (\autoref{tab:arcstable},\autoref{fig:model}), belonging to 7 unique sources (the sources labeled 3 and 5 may be associated with the same galaxy). 
At the time of the computation of the model, none of the lensed sources had spectroscopic redshifts. We obtained photometric redshifts based on the extensive \hst\ imaging, using the BPZ \citep{bpz2000} algorithm \citep[see][]{Cerny2018, coe2019, Salmon2020}.

\begin{deluxetable}{lll} 
\tablecolumns{3} 
\tablecaption{List of lensing constraints \label{tab:arcstable}} 
\tablehead{\colhead{ID} &
            \colhead{R.A. [deg]}    & 
            \colhead{Decl. [deg]}    \\[-8pt]
            \colhead{} &
            \colhead{J2000}     & 
            \colhead{J2000}     }
\startdata 
\hline
1.1 & 110.80504 & $-$73.454536 \\  
1.2 & 110.80681 & $-$73.458358 \\ 
1.3 & 110.81306 & $-$73.448694 \\
Source 2: & &\\
2.1 & 110.84047 & $-$73.450975\\
2.2 & 110.84270 & $-$73.454747\\
2.3 & 110.83876 & $-$73.458692\\
Source 3: & &\\
3.1 & 110.82491 & $-$73.459581\\
3.2 & 110.83152 & $-$73.455153\\
3.3 & 110.83019 & $-$73.448453\\
3.4 & 110.82297 & $-$73.454786\\ 
Source 4: & &\\
4.1 & 110.82364 & $-$73.451800\\
4.2 & 110.82210 & $-$73.452686\\
4.3 & 110.82067 & $-$73.460128\\
Source 5: & &\\
5.1 & 110.82522 & $-$73.459686\\
5.2 & 110.83176 & $-$73.455108\\
5.3 & 110.83027 & $-$73.448542\\
5.4 & 110.82316 & $-$73.454736\\
Source 6: & &\\
6.1 & 110.83212 & $-$73.454375\\
6.2 & 110.82170 & $-$73.454108\\
6.3 & 110.82948 & $-$73.448914\\
6.4 & 110.82289 & $-$73.461631\\
Source 7: & &\\
7.1 & 110.83560 & $-$73.451735\\
7.2 & 110.83652 & $-$73.453005\\
Source 8: & &\\
8.1 & 110.83842 & $-$73.451014\\
8.3 & 110.83610 & $-$73.458784\\
\enddata 
\tablecomments{The coordinates match the WCS solution of the RELICS data reduction version \texttt{v1}, which is available on MAST. 
}
\end{deluxetable}

\subsection{Lens model components}
We modeled the cluster using the public software \lenstool\ \citep{jullo07}. This algorithm uses MCMC formalism to explore the parameter space and identify the best-fit set of parameters, which minimize the scatter between the observed and predicted lensing evidence. The cluster component was represented by a parametric pseudo-isothermal mass distribution halo (PIEMD, a.k.a dPIE; \citealt{eliasdottir07}), with parameters $x$, $y$, $e$, $\theta$, $r_{core}$, $r_{cut}$, and $\sigma$. All the parameters were allowed to vary within broad priors, except for $r_{cut}=1000$ kpc. Another PIEMD halo represented the brightest cluster galaxy (BCG), with $x$, $y$, $e$ and $\theta$ fixed to the observed values as measured with Source Extractor \citep{Bertin1996} in the F814W image, and the others allowed to vary. We selected cluster-member galaxies based on their color in a F606W-F814W vs. F814W diagram using the red sequence technique \citep{gladdersyee2000}. To measure magnitudes and colors, we used Source Extractor \citep{Bertin1996} in dual-image mode with the F814W band used for reference and photometry. Stars were identified and removed from the catalog based on their location in a \texttt{MU\_MAX} vs \texttt{MAG\_AUTO} diagram. Galaxies that were selected as cluster members were included as PIEMD halos in the model. The positional parameters of the galaxies were fixed to their catalog values, whereas $r_{core}$, $r_{cut}$ and  $\sigma$ were determined by \lenstool\ based on the scaling relations that are described in \citet{limousin05}, with pivot parameters \texttt{mag0}=19.12 mag and \texttt{corekpc}=0.15 kpc; the scaling relation parameters \texttt{sigma} and \texttt{cutkpc} were optimized by the model. The model used a total of 145 halos, of which two were individually optimized. 

We fixed the redshift of Source~1 to the its photometric redshift, $\zphot=2.2$. The redshifts of all the other lensed sources were entered as free parameters with broad flat priors ($0.5<z<5$). 
The broad priors were used in order to not be affected by possible catastrophic photo-$z$ outliers. Some photo-$z$ measurements exhibited large uncertainties, or gave inconsistent results for multiple images of the same source. This was in part due to contamination from other sources (e.g., images of source 3/5 are projected near a bright star). 
Other than fixing one redshift to the most secure photo-$z$, we only used the photo-$z$ information statistically, to check that overall the lens model predictions are consistent with the photometric redshifts (see \citealt{Cerny2018} for a description of this approach). 
The diagnostic plot of model-$z$ vs. photo-$z$ is shown in the bottom-right panel of \autoref{fig:model}. 

Including the free redshifts, this model has a total of 18 free parameters (6 for the cluster halo, 3 for the BCG halo, 2 for cluster member galaxies scaling, and 7 free redshifts), and 34 constraints (25 images from 8 sources).

\section{results}
The observed images of lensed sources were well-reproduced by the lens model, with a typical image plane scatter per system of $<0\farcs5$. A somewhat higher image plane scatter for System 2 drove the overall image-plane RMS to $0\farcs58$. 
\autoref{fig:model} shows the lensing constraints and the best-fit critical curve for a source at $z=3$, overplotted on the \hst\ imaging. 
The bottom-right panel of \autoref{fig:model} shows a comparison of the model-predicted redshift and the photometric redshifts measured by RELICS. As noted above, this plot was used as a diagnostic tool to assess the model, in lieu of spectroscopic redshifts. We found that overall the model did not appear to systematically over-predict or under-predict the photometric redshifts.

Spectroscopic redshifts were not available at the time this model was computed, but were recently published by several authors \citep{mahler2022,Golubchik2022}.
We compare the spectroscopic redshifts from \citet{mahler2022} to the posterior distributions of the free redshifts of systems 2--8 (the redshift of System 1 was fixed to its photo-$z$, $\zphot=2.2$, which does not have a spectroscopic redshift). The best-fit redshift of each system is indicated in orange, and the spectroscopic redshift in blue.  We find that the model correctly predicted the redshift(s) of system(s) 3/5, but underpredicted the redshifts of systems 2, 4, and 8 by $\Delta z / (1+z) = 0.065, 0.027, 0.062$, respectively. 

RELICS made the following lensing products publicly available through MAST: shear ($\gamma$), convergence ($\kappa$), lensing potential ($\psi$), deflection in the $x$ and $y$ direction ($\alpha_x$,$\alpha_y$), and magnification maps ($\mu(z)$) for several redshifts. With the exception of magnification, the files are scaled to effectively $D_{ls}/D_{s}=1$, where $D_{ls}$, $D_{s}$ are the angular diameter distances from the lens to the source and from the observer to the source, respectively. They can be re-scaled to any source redshift by multiplying by the relevant $D_{ls}/D_{s}$. We also made available a set of 100 files of each of the above, sampled from the MCMC chain, which can be used to estimate the statistical uncertainties related to the lens modeling process. 

This paper complements the public models and provides context for current works that aim to take full advantage of the new \JWST\ data, and have either already used the public models, or wish to 
compare a \JWST-based lensing analyses to the pre-\JWST\ models. 

\acknowledgements
Based on observations with the NASA/ESA \HST, obtained at STScI, which is operated by AURA under NASA contract NAS5-26555, associated with the RELICS Treasury Program GO-14096, and with programs GO-12166 and GO-12884.
The data were obtained from Mikulski Archive for Space Telescopes (MAST). 
Support for GO-14096 was provided through a grant from the STScI under NASA contract NAS5-26555.

\vspace{5mm}
\facilities{HST(ACS), HST(WFC3), HST(MAST)}

\software{\lenstool,  \citep{jullo07}
          Source Extractor \citep{Bertin1996}
          MATLAB Astronomy and Astrophysics Toolbox \citep[MAAT][]{Ofek2014}
          }

\bibliography{smacs}{}
\bibliographystyle{aasjournal}

\end{document}